\documentclass[12pt,epsf,epsfig]{article}
\usepackage{graphicx}
\usepackage{amsmath,latexsym}

\begin{document}

\begin{center}
{\huge{A Born-Infeld Scalar and a Dynamical Violation of the Scale Invariance from the Modified Measure Action.}} \\
\end{center}

\begin{center}
T.O. Vulfs \textsuperscript{1}, E.I. Guendelman \textsuperscript{1,2,3} \\
\end{center}

\begin{center}
\textsuperscript{1} Department of Physics, Ben-Gurion University of the Negev, Beer-Sheva, Israel \\
\end{center}

\begin{center}
\textsuperscript{2} Frankfurt Institute for Advanced Studies, Giersch Science Center, Campus Riedberg, Frankfurt am Main, Germany \\
\end{center}

\begin{center}
\textsuperscript{3} Bahamas Advanced Study Institute and Conferences, 4A Ocean Heights, Hill View Circle, Stella Maris, Long Island, The Bahamas \\
\end{center}

E-mail: vulfs@post.bgu.ac.il, guendel@bgu.ac.il

\abstract
Starting with a simple two scalar field system coupled to a modified measure that is independent of the metric, we, first, find a Born-Infeld dynamics sector  of the theory for a scalar field and second, show that the initial scale invariance of the action is dynamically broken and leads to a scale charge nonconservation, although there is still a conserved dilatation current.

\section{Introduction}

Every physical quantity is defined by its transformation properties. A scalar field, $\phi$, being a single real function of spacetime behaves as a scalar under Lorentz transformations: $\phi(x) \rightarrow \phi'(x')$. Its dynamics is determined by the kinetic and potential energy densities. In our paper we show, first, how this simplest scalar field can be transformed to a Born-Infeld scalar and second, how the dynamically broken scale invariance leads to the nonconservation of a scale charge. It all becomes possible when we use a modified measure in the action instead of a standard one. \\

The notion of a measure is usually associated with the theories of gravity. There $\sqrt{-g}$, where $g$ is the determinant of the metric, is included to the action, i.e. $S = \int L \sqrt{-g} dx$, to make the volume element invariant. However, such choice for the measure is not unique. The only requirement that it must be a density under diffeomorphic transformations can be fulfilled in other ways. Our new measure is

\begin{equation} \label{eq:0}
\Phi = \epsilon^{\mu_1\mu_2\ldots\mu_D}\epsilon_{a_1a_2\ldots a_D}\partial_{\mu_1}\varphi_{a_1}\ldots\partial_{\mu_D}\varphi_{a_D},
\end{equation}

where $\epsilon^{\mu_1\mu_2\ldots\mu_D}$ and $\epsilon_{a_1a_2\ldots a_D}$ are Levi-Civita symbols and $\varphi_{a_1}\ldots\varphi_{a_D}$ are additional scalar fields that have nothing to do with the original $\phi$. This modified measure is already applied in gravity as first proposed in \cite{d,e}. Different from (\ref{eq:0}), the Galileon measure is considered in String Theory in \cite{e1,e2}.  \\

We assume that our spacetime is two-dimensional to avoid any unnecessary complications. Then

\begin{equation}
\Phi = \epsilon^{\mu\nu}\epsilon_{ab} \partial_\mu \varphi_a \partial_{\nu} \varphi_b.
\end{equation}

We choose this particular realization for $\Phi$ because it is appropriate for our goals. We are able to do it because

\begin{equation}
\Phi \rightarrow \det(\frac{\partial x^{\mu'}}{\partial x^{\mu}})^{-1} \Phi, \quad d^2 x \rightarrow \det(\frac{\partial x^{\mu'}}{\partial x^{\mu}}) d^2 x.
\end{equation}

Therefore,

\begin{equation}
\Phi d^2 x \rightarrow \Phi d^2 x.
\end{equation}

The general form of the action is

\begin{equation} \label{eq:5}
S = \int \Phi L d^D x.
\end{equation}

To avoid a confusion in the terminology: the lagrangian is $\Phi L$, let's call it $L_{full}$ and by $L$ we mean the part of $L_{full}$ without the measure $\Phi$. Notice that when the measure appears only linearly, as in (\ref{eq:5}) and the measure fields $\varphi_a$ do not enter in $L$, there is an infinite dimensional symmetry

\begin{equation}
\varphi_a \rightarrow \varphi_a + f_a(L),
\end{equation}

as has been discussed in \cite{d}. \\

As the background is clear, let's check what exactly the goals are. \\

The essence of the Born-Infeld theory is the requirement of finitness of a physical quantities. Originated as a specific theory of nonlinear electrodynamics in \cite{c}, it put limitations on the self-energy of a point charge. Later it reappeared in string theory to describe the electromagnetic fields on the world-volumes of D-branes as it guarantees that the energy of the string is finite in \cite{g,f}. Recently, to bring limits on scalar fields in cosmology, the Born-Infeld scalar was considered in \cite{h,i}. This integration was developed later in \cite{i1,i2,i3,i4,i5,i6,i7,i8}. Our first aim is a naturally arising restraints on our scalar field. \\

The essence of the scale invariance is the requirement that physics must be the same at all scales, i.e. the system must be invariant under the global scale transformations ($\omega$ is a constant):

\begin{equation} \label{eq:11}
g_{\mu\nu} \rightarrow \omega g_{\mu\nu}.
\end{equation}

However, the physical universe definitely does not have such property, and different scales show different behavior. Then to approach reality the scale invariance must be broken. Our second aim is a naturally arising dynamical violation of the scale invariance. In this paper we are going to work with a fixed background metric, so the transformation (\ref{eq:11}) will not be used, instead the fields will transform and a volume element independent of the metric will be allowed to transform. \\

Moreover, according to Noether's theorem, the symmetries and conservation laws are tightly connected. Then our third aim is to show that despite being a symmetry, the scale invariance does not lead to the conservation of the scale charge. \\

The anomalous infrared behavior of the conserved chiral current in the presence of instantons was discussed in \cite{b}. The conclusion was made that in this case there was no conserved $U(1)$ charge and Goldstone's theorem therefore failed, solving the $U(1)$ problem in QCD. The case of global scale invariance in the presence of a modified measure was considered in \cite{a,a1,a2,a3,a4,a5,a6,a7,a8} and the dilatation currents were calculated in a special model in \cite{a0}, where the current was shown to be singular in the infrared. Here also, the resulting scale current produces a nonzero flux of the dilatation current to infinity, so once again, although there is a conserved current, there is no conserved scalar charge. \\

Section 2 is devoted to the preparations for the later sections: the guiding principles are considered in more details and the lagrangian is provided. In Section 3 we arrive at the Born-Infeld scalar dynamics. In Section 4 we show how the asymptotic behavior of the conserved current leads to the nonconservation of a scale charge. The conclusions are given in Section 5. \\

\section{General Considerations}

The modified measure results in the dynamical violation of the scale invariance. To see that, we consider the variation of (\ref{eq:5}) with respect to $\varphi_a$:

\begin{equation}
A^{\mu_1}_{a_1} \partial_{\mu_1} L = 0,
\end{equation}

where $A^{\mu_1}_{a_1} = \epsilon^{\mu_1\mu_2\ldots\mu_D}\epsilon_{a_1a_2\ldots a_D}\partial_{\mu_2}\varphi_{a_2}\ldots\partial_{\mu_D}\varphi_{a_D}$ and we assume that $L$ is independent of $\varphi_a$'s. \\

If $\det(A^{\mu_i}_{a_j}) \sim \Phi$ is non-trivial, then the solution is

\begin{equation} \label{eq:4}
L = M = constant.
\end{equation}

The appearance of the constant in (\ref{eq:4}) breaks the scale invariance. \\

To break the scale invariance in a consequence, the action must be scale invariant initially. Then

\begin{equation}
S = \frac12 \int \Phi \partial_{\mu} \phi \partial_{\nu}\phi g^{\mu\nu} d^2 x.
\end{equation}

Without loss of generality, we assume that the scalar field possesses only the kinetic energy. \\

The theory has the scale invariance with the following choice of the rescaling of the fields:

\begin{equation} \label{eq:20}
\phi \rightarrow \lambda^{-\frac12} \phi, \quad \varphi_a \rightarrow \lambda^{\frac12} \varphi_a,
\end{equation}

where $\lambda$ is the rescaling parameter that applies to the scalar fields and the measure only (and the metric remains invariant). \\

However, it turns out that this model is not able to bring the enviable results. We must add one more scalar to the lagrangian. \\

The final lagrangian is

\begin{equation}
L = \frac12 (\partial_{\mu} \phi_1\partial_{\nu}\phi_1 g^{\mu\nu} + \partial_{\mu} \phi_2\partial_{\nu}\phi_2 g^{\mu\nu}),
\end{equation}

where $\phi_1$ is the former scalar field $\phi$ and $\phi_2$ is the supplemented one. \\

So that the final action is

\begin{equation} \label{eq:111}
S = \frac12 \int \Phi (\partial_{\mu} \phi_1\partial_{\nu}\phi_1 g^{\mu\nu} + \partial_{\mu} \phi_2\partial_{\nu}\phi_2 g^{\mu\nu}) d^2 x,
\end{equation}

where for $S$ to be scale invariant we choose the rescaling of the additional field as

\begin{equation} \label{eq:21}
\phi_2 \rightarrow \lambda^{-\frac12} \phi_2.
\end{equation}

So at that level the dynamics of the initial scalar field $\phi$ is defined by the action (\ref{eq:111}). To achieve our aims we added to that action three more scalar fields: $\phi_2$ is physically equivalent to $\phi$ and enter the lagrangian in the same footing, $\varphi_{a}$ $(a=1,2)$ are the base for the newly constructed modified measure in 2D. In the following sections we show how such complexity leads to the solutions. \\

A step further is to obtain the equations of motion, i.e. the variations of $S$ with respect to the dynamical variables. \\

For simplicity we consider flat Minkowski 2D spacetime so that

\begin{equation}
g_{\mu\nu} = \eta_{\mu\nu},
\end{equation}

with the signature $(- +)$. \\

\section{The Appearance of a Born-Infeld Scalar Sector.}

The dynamical variables of (\ref{eq:111}) are $\phi_1$, $\phi_2$, $\varphi_a$. \\

The variation with respect to $\varphi_a$ is

\begin{equation}
\epsilon^{\mu\nu} \epsilon_{ab}\partial_{\nu} \varphi_{b} \partial_{\mu} L =  0.
\end{equation}

If $\Phi$ is non-trivial then $\epsilon^{\mu\nu} \epsilon_{ab}\partial_{\nu} \phi_{b}$ is non-trivial. Then we can obtain

\begin{equation} \label{eq:7}
L = \frac12 (\partial_{\mu} \phi_1\partial_{\nu}\phi_1 g^{\mu\nu} + \partial_{\mu} \phi_2\partial_{\nu}\phi_2 g^{\mu\nu}) = M.
\end{equation}

Note that we can see (even before studying the Born-Infeld scalar sector) that for static case the gradients of the two scalar fields are bounded by $\sqrt{2M}$. \\

(\ref{eq:7}) is rewritten to give

\begin{equation} \label{eq:10}
(\partial_{\mu} \phi_1)^2 + (\partial_{\mu} \phi_2)^2 = 2M.
\end{equation}

It is interesting to notice that this is a kind of "nonlinear gradient $\sigma$ model". \\

The variation with respect to $\phi_2$ is

\begin{equation} \label{eq:99}
\partial_{\mu} (\Phi \partial^{\mu} \phi_2) = 0.
\end{equation}

We assume that $\phi_2$ depends only on the spatial coordinate $x$, $\phi_2 = \phi_2(x)$. Then (\ref{eq:99}) becomes

\begin{equation}
\partial_1 (\Phi \partial_1 \phi_2) = 0,
\end{equation}

which can be integrated to give

\begin{equation}
\Phi \partial_1 \phi_2 = J = constant.
\end{equation}

We observe that the action has the additional shift symmetry:

\begin{equation} \label{eq:9}
\phi_2 \rightarrow \phi_2 + c_2,
\end{equation}

where $c_2$ is a constant. \\

This symmetry leads to the conservation law. Then $J$ has the interpretation of a constant current flowing in the $x$-direction. Then

\begin{equation}
\partial_1 \phi_2 = \frac{J}{\Phi}.
\end{equation}

Inserting this into (\ref{eq:7}), we get

\begin{equation}
\partial_{\mu} \phi_1 \partial^{\mu} \phi_1 + \frac{J^2}{\Phi^2} = 2M,
\end{equation}

which can be used to solve for the measure $\Phi$, giving

\begin{equation} \label{eq:22}
\Phi = \frac{J}{\sqrt{2M - \partial_{\mu}\phi_1\partial^{\mu}\phi_1}}.
\end{equation}

The variation with respect to $\phi_1$ is

\begin{equation} \label{eq:23}
\partial_{\mu} (\Phi \partial^{\mu} \phi_1) = 0.
\end{equation}

Making the same assumptions as for the $\phi_2$, namely $\phi_1 = \phi_1(x)$ and

\begin{equation} \label{eq:8}
\phi_1 \rightarrow \phi_1 + c_1,
\end{equation}

where $c_1$ is a constant, we obtain

\begin{equation}
\partial_{\mu}(\Phi \partial^{\mu} \phi_1) = 0.
\end{equation}

Inserting (\ref{eq:22}) into (\ref{eq:23}), we get the Born-Infeld scalar equation (for $M >0$):

\begin{equation} \label{eq:100}
J \partial_{\mu}(\frac{\partial^{\mu}\phi_1}{\sqrt{2M - \partial_{\mu}\phi_1\partial^{\mu}\phi_1}}) = 0,
\end{equation}

which is also obtained from the effective Born-Infeld action for this kind of solutions.

\begin{equation} \label{eq:101}
S_{eff} = \int \sqrt{2M - \partial_{\mu}\phi_1\partial^{\mu}\phi_1} d^2 x.
\end{equation}

The dynamics of $\phi_1$ defined by the equation (\ref{eq:100}) is the same as the dynamics of $\phi_1$ derived from the variation of (\ref{eq:101}). This means that $\partial_{\mu}\phi_1\partial^{\mu}\phi_1$ is bounded in this sector of the theory (the Born-Infeld scalar sector) Notice, however, that we are now considering only a sector of the theory. \\

\section{The Breaking of Charge Conservation.}

The action (\ref{eq:111}) is invariant under the scale transformations (\ref{eq:20}) and (\ref{eq:21}). By the Noether's theorem a conservation quantity must appear, namely, the scale charge, $Q$. Then the continuity equation must be satisfied: \\

\begin{equation}
\frac{\partial \rho}{\partial t} + \nabla j = 0,
\end{equation}

where $\rho$ is a density of $Q$ and $j$ is the flux of $Q$. \\

By the the integration

\begin{equation}
\int_{x_1}^{x^2} \frac{\partial \rho}{\partial t} d^2 x + \int_{x_1}^{x^2} \frac{\partial j^1}{\partial x} d^2 x = 0
\end{equation}

we obtain

\begin{equation}
\frac{d Q}{dt} + j^1(x_2) - j^1(x_1)=0.
\end{equation}

Therefore, the conservation of the total charge requires

\begin{equation}
j^1(x_1 \rightarrow -\infty)-j^1(x_2 \rightarrow +\infty) = 0.
\end{equation}

However, it does not always happen. Our case is one of the exceptions. \\

The conserved current is given by \\

\begin{equation} \label{eq:1}
j^{\mu} = \frac{\partial L_{full}}{\partial(\partial_{\mu}\varphi_a)}\delta \varphi_a + \frac{\partial L_{full}}{\partial(\partial_{\mu}\phi_1)}\delta \phi_1 + \frac{\partial L_{full}}{\partial(\partial_{\mu}\phi_2)}\delta \phi_2.
\end{equation}

We consider a scale transformations infinitesimally closed to the identity: $\lambda = (1 + \theta)$, so that (\ref{eq:20}) and (\ref{eq:21}) turn into

\begin{equation}
\varphi_a \rightarrow (1 + \theta)^{\frac12}\varphi_a \simeq \varphi_a + \frac{\theta}{2}\varphi_a,
\end{equation}

\begin{equation}
\phi_1 \rightarrow (1 + \theta)^{-\frac12}\phi_1 \simeq \phi_1 - \frac{\theta}{2}\phi_1,
\end{equation}

\begin{equation}
\phi_2 \rightarrow (1 + \theta)^{-\frac12}\phi_2 \simeq \phi_2 - \frac{\theta}{2}\phi_2.
\end{equation}

Therefore,

\begin{equation}
\delta\varphi_a = \frac{\theta}{2}\varphi_a, \quad \delta\phi_1 = -\frac{\theta}{2}\phi_1, \quad \delta\phi_1 = -\frac{\theta}{2}\phi_2.
\end{equation}

Then (\ref{eq:1}) becomes

\begin{equation}
j^{\mu} = M\frac{\theta}{2}\epsilon^{\mu\nu}\epsilon_{ab}\varphi_a\partial_{\nu}\varphi_b - \frac{\theta}{2}\Phi\partial^{\mu}\phi_1 - \frac{\theta}{2}\Phi\partial^{\mu}\phi_2.
\end{equation}

Let's go back to (\ref{eq:100}) and find the static solutions ($\partial_0 \phi_1 = 0$):

\begin{equation}
\partial_1 (\frac{\partial_1 \phi_1}{\sqrt{2M - (\partial_1 \phi_1)^2}}) = 0.
\end{equation}

By integration we get

\begin{equation}
\frac{\partial_1 \phi_1}{\sqrt{2M - (\partial_1 \phi_1)^2}} = c_3,
\end{equation}

where $c_3$ is a constant. \\

Then

\begin{equation}
\phi_1 = \frac{\sqrt{2M} |c_3|}{\sqrt{1 + |c_3|^2}} (x_2-x_1).
\end{equation}

We have done all the calculations for $\phi_1$. However, the same is relevant for $\phi_2$. So that

\begin{equation}
\phi_2 = \frac{\sqrt{2M} |c_4|}{\sqrt{1 + |c_4|^2}} (x_2-x_1),
\end{equation}

where $c_4$ is a constant. \\

Inserting this solution to (\ref{eq:22}), we obtain

\begin{equation}
\Phi = J \sqrt{\frac{1 + c_3^2}{(1 - 2M)c_3^2 + 1}}.
\end{equation}

Then we see

\begin{equation}
\Phi = \Phi_0 = constant.
\end{equation}

It is satisfied for

\begin{equation}
\varphi_1 = c_5 t, \quad \varphi_2 = c_6 x,
\end{equation}

where $c_5$ and $c_6$ are constants. \\

Then indeed

\begin{equation}
\Phi = c_5 c_6.
\end{equation}

Now by inserting the solutions for $\varphi_a$, $\phi_1$ and $\phi_2$ into (\ref{eq:1}), we obtain for $j^1$

\begin{equation}
j^1 = \frac{\theta}{2}M c_5 c_6 x - \frac{\theta}{2}c_5 c_6\frac{\sqrt{2M}|c_3|}{\sqrt{1+|c_3|^2}} - \frac{\theta}{2}c_5 c_6\frac{\sqrt{2M}|c_4|}{\sqrt{1+|c_4|^2}}.
\end{equation}

We see that $j^1$ is a constant plus a term proportional to $x$ and therefore, $j^1(\infty)-j^1(-\infty) \ne 0$ and in fact diverges. Therefore, $Q$ is not conserved. \\

Let's calculate $j^0$ explicitly.

\begin{equation}
j^0 = -\frac{\theta}{2}Mc_5 c_6 t.
\end{equation}

Then we see that

\begin{equation}
Q = \int_{x_1}^{x^2} j^0 dx = -(x_2 - x_1)\frac{\theta}{2}M c_5 c_6 t.
\end{equation}

We checked that $Q$ is not conserved. \\

\section{Conclusions}

In this paper we start with the scalar field, surround it with three supplementary scalar fields and investigate the resulting action, (\ref{eq:111}). One scalar field is physically equivalent to the former scalar field. However, the new measure of integration is constructed from the other two scalars. The source of the following findings is this modified measure. First, we show that the gradient of this initial scalar field is finite and in particular there is a sector which can be presented in the form of the Born-Infeld scalar. Second, the initial action is scale invariant, however, the invariance gets spontaneously broken. In addition to having spontaneous symmetry breaking, our physical system serves as an example of a system with the symmetry that does not lead to the conserved charge. \\

A consideration of a scale invariance in cosmology started in \cite{v1,v2} and was continued in \cite{a10,a11}. In this case when the scale symmetry is spontaneously broken, there is a conserved current and since no singular behavior of the conserved current is obtained, so there is a conserved scale charge and the Goldstone theorem holds. \\

Note that in our case of a scalar field there remains a massless field which is a Goldstone Boson of the shift symmetry $(\phi_1 \rightarrow \phi_1 + constant)$, not of the scale symmetry, because for the scale symmetry that theorem cannot be applied since the dilatation charge is not conserved. \\

\textbf{Acknowledgments}
TV acknowledges support by the Ministry of Aliyah and Integration (IL) and the Israel Science Foundation. EG is supported by the Foundational Questions Institute and COST actions CA15117, CA16104, CA18108. We thank Emil Nissimov and Svetlana Pacheva for interesting discussions.


\begin{thebibliography}{28}
\bibitem{d}
E.I. Guendelman, A.B. Kaganovich, Phys.Rev.D53, 7020, (1996), [gr-qc/9605026]
\bibitem{e}
E.I. Guendelman, A.B. Kaganovich, Phys.Rev.D55, 5970, (1997), [gr-qc/9611046]
\bibitem{e1}
T.O. Vulfs, E.I. Guendelman, Mod.Phys.Lett. A32, no.38, 1750211, (2017), [arXiv:1708.00458]
\bibitem{e2}
T.O. Vulfs, E.I. Guendelman, Annals Phys. 398 (2018) 138-145, (2018), [arXiv:1709.01326]
\bibitem{c}
M. Born, L. Infeld, Proc.Roy.Soc.Lond. A144, no.852, 425-451, (1934)
\bibitem{g}
C.G. Callan, J.M. Maldacena, Nucl.Phys. B513, 198-212, (1998), [hep-th/9708147]
\bibitem{f}
G.W. Gibbons, Nucl.Phys. B514, 603-639, (1998), [hep-th/9709027]
\bibitem{h}
J.A. Feigenbaum, P.G.O. Freund, M. Pigli, Phys.Rev. D57, 4738-4744, (1998), [hep-th/9709196]
\bibitem{i}
S. Deser, G.W. Gibbons, Class.Quant.Grav. 15, L35-L39, (1998), [hep-th/9803049]
\bibitem{i1}
G.N. Felder, L. Kofman, , A. Starobinsky, JHEP 0209, 026, (2002), [hep-th/0208019]
\bibitem{i2}
D.N. Vollick, Gen.Rel.Grav. 35, 1511-1516, (2003), [hep-th/0102187]
\bibitem{i3}
Jian-gang Hao, Xin-zhou Li, Phys.Rev. D68, 043501, (2003), [hep-th/0305207]
\bibitem{i4}
Dan N. Vollick, Phys.Rev. D72, 084026, (2005), [gr-qc/0506091]
\bibitem{i5}
W. Fang, H.Q. Lu, Z.G. Huang, K.F. Zhang, Int.J.Mod.Phys. D15, 199-214, (2006), [hep-th/0409080]
\bibitem{i6}
S. Jana, S. Kar, Phys.Rev. D94, no.6, 064016, (2016), [arXiv:1605.00820]
\bibitem{i7}
V.I. Afonso, G.J. Olmo, D. Rubiera-Garcia, JCAP 1708, no.08, 031, (2017), [arXiv:1705.01065]
\bibitem{i8}
S. Jana, S. Kar, Phys.Rev. D96, no.2, 024050, (2017), [arXiv:1706.03209]
\bibitem{b}
G. 't Hooft, Phys.Rept. 142, 357-387, (1986)
\bibitem{a}
E.I. Guendelman, Class.Quant.Grav. 17, 361-372, (2000), [gr-qc/9906025]
\bibitem{a1}
E.I. Guendelman, R. Herrera, P. Labrana, E. Nissimov, S. Pacheva, Gen.Rel.Grav. 47, no.2, 10, (2015), [arXiv:1408.5344]
\bibitem{a2}
E.I. Guendelman, H. Nishino, S. Rajpoot, Phys.Lett. B732, 156-160, (2014), [arXiv:1403.4199]
\bibitem{a3}
E.I. Guendelman, E. Nissimov, S. Pacheva, M. Vasihoun, Bulg.J.Phys. 40, 121-126, (2013), [arXiv:1310.2772]
\bibitem{a4}
S. del Campo, E.I. Guendelman, A.B. Kaganovich, R. Herrera, P. Labrana, Phys.Lett. B699, 211-216, (2011), [arXiv:1105.0651]
\bibitem{a5}
S. del Campo, E.I. Guendelman, R. Herrera, P. Labrana, JCAP 1006, 026, (2010), [arXiv:1006.5734]
\bibitem{a6}
E.I. Guendelman, A.B. Kaganovich, Annals Phys. 323, 866-882, (2008), [arXiv:0704.1998]
\bibitem{a7}
E.I. Guendelman, A.B. Kaganovich, Phys.Rev. D75, 083505, (2007), [gr-qc/0607111]
\bibitem{a8}
E.I. Guendelman, O. Katz, Class.Quant.Grav. 20, 1715-1728, (2003), [gr-qc/0211095]
\bibitem{a0}
E.I. Guendelman, Mod.Phys.Lett. A14, 1043-1052, (1999), [gr-qc/9901017]
\bibitem{v1}
J. Garcia-Bellido, J. Rubio, M. Shaposhnikov, D. Zenhausern, Phys.Rev. D84, 123504, (2011), [arXiv:1107.2163]
\bibitem{v2}
F. Bezrukov, G.K. Karananas, J. Rubio, M. Shaposhnikov, Phys.Rev. D87 (2013) no.9, 096001, (2013), [arXiv:1212.4148]
\bibitem{a10}
P.G. Ferreira, C.T. Hill, G.G. Ross, Phys.Rev. D98, no.11, 116012, (2018),  [arXiv:1801.07676]
\bibitem{a11}
P. G. Ferreira, C.T. Hill, J. Noller, G.G. Ross, Phys.Rev. D97, no.12, 123516, (2018), [arXiv:1802.06069]

\end{thebibliography}
\end{document}